\documentclass[prb,twocolumn,superscriptaddress,showpacs,floatfix]{revtex4}
\usepackage{graphicx}% Include figure files
\usepackage{epsfig}% Include figure files
\usepackage{amssymb,amsmath,amsfonts,hyperref}
\usepackage{wasysym}
\usepackage{latexsym}
\usepackage{color}

\newcommand{\be}{\begin{eqnarray}}
\newcommand{\ee}{\end{eqnarray}}

\begin{document}
%\preprint{APS/123-QED}
%draft
%\twocolumn[\hsize\textwidth\columnwidth\hsize\csname @twocolumnfalse\endcsname
\title{Antiferromagnetic order in systems with doublet $S_{\rm tot}=1/2$ ground states}
\author{Sambuddha Sanyal}
\affiliation{Department of Theoretical Physics, Tata Institute of Fundamental Research, Mumbai 400005, India.}
\author{Argha Banerjee}
\affiliation{Department of Theoretical Physics, Tata Institute of Fundamental Research, Mumbai 400005, India.}
\author{Kedar Damle}
\affiliation{Department of Theoretical Physics, Tata Institute of Fundamental Research, Mumbai 400005, India.}
\author{Anders W. Sandvik}
\affiliation{Department of Physics, Boston University, 590 Commonwealth Avenue,
Boston, Massachussetts 02215, USA.}

\begin{abstract}
We use projector Quantum Monte-Carlo methods
to study the $S_{\rm tot}=1/2$ doublet ground states of two dimensional $S=1/2$ antiferromagnets 
on a $L \times L$ square lattice with an odd number of sites $N_{\rm tot}=L^2$. We compute the ground 
state spin texture $\Phi^z(\vec{r}) = \langle S^z(\vec{r})\rangle_{\uparrow}$ in 
$|G \rangle_{\uparrow}$, the $S^z_{\rm tot}=1/2$ component of
this doublet, and investigate the relationship between $n^z$, the
thermodynamic
limit of the staggered component of this ground state spin texture,
and $m$, the thermodynamic limit of the magnitude of the staggered magnetization vector of the same system in the singlet
ground state that obtains for even $N_{\rm tot}$. $n^z$ and $m$ would
have been equal if the non-zero
value of $S^z_{\rm tot}$ in $|G \rangle_{\uparrow}$ caused the direction of
the staggered magnetization vector to be fully pinned in the thermodynamic
limit. By studying several
different deformations of the square lattice Heisenberg antiferromagnet,
we establish that this is not the case.
For the sizeable range of $m$ accessed in our numerics, we find a univeral relationship between the two, that is independent of the microscopic details of the lattice level Hamiltonian
and can be well approximated by a polynomial interpolation formula: $n^z \approx (\frac{1}{3} - \frac{a}{2} -\frac{b}{4}) m + am^2+bm^3$, with $a \approx 0.288$ and $b\approx -0.306$.
We also find that the full spin texture $\Phi^z(\vec{r})$ is itself dominated
by Fourier modes near the antiferromagnetic wavevector
in a universal way. On the analytical side, we
explore this question using spin-wave theory, a simple
mean field model written in terms of the total spin of each sublattice,
and a rotor model for the dynamics of $\vec{n}$.
We find that spin-wave theory reproduces this universality of $\Phi^z(\vec{r})$ and gives $n^z = (1-\alpha -\beta/S)m + (\alpha/S)m^2 +{\mathcal O}(S^{-2})$ with $\alpha \approx 0.013$ and $\beta \approx 1.003$ for spin-$S$ antiferromagnets, while the sublattice-spin mean field theory and the rotor model both give $n^z = \frac{1}{3} m$ for $S=1/2$ antiferromagnets. We argue that this latter
relationship  becomes asymptotically exact in the limit
of infinitely long-range {\em unfrustrated} exchange interactions.

\end{abstract}

\vskip2pc

\maketitle

\section{Introduction}
Computational studies of strongly correlated
systems necessarily involve an extrapolation to the thermodynamic limit from a sequence of finite
sizes at which calculations are feasible. Understanding,\cite{Neuberger_Ziman} and at times reducing,\cite{Chernyshev_White} these 
finite-size corrections to the thermodynamic limit is thus an
important aspect of any such calculation.
For instance, the best estimates of $m$, the magnitude of the ground state
N\'eel order parameter in the thermodynamic limit of the two-dimensional $S=1/2$
square lattice Heisenberg antiferromagnet rely on a sequence of 
$L_x \times L_y$ systems with even length $L_x$ ($L_y$) in the $x$
($y$) direction and periodic boundary conditions in both directions.\cite{Sandvik_prb97,Beard_Wiese_prb}
Other studies suggest \cite{Chernyshev_White} that it is some times advantageous
to use ``cylindrical'' samples with periodic boundary conditions in one direction and pinned boundary conditions 
in the other direction, whereby spins are held fixed by the use of pinning fields on one pair of edges---this choice
also allows for a very accurate determination of ground-state parameters such as $m$ for specific 
values\cite{Chernyshev_White} of the aspect ratio $L_y/L_x$.

All these approaches focus on systems with an {\em even} number of
spin-half variables; this choice allows the ground-state of the finite
system to lie in the singlet sector favoured by unfrustrated antiferromagnetic interactions.\cite{Lieb_Mattis}
Although not commonly used, another choice is certainly
possible: Namely, one could in principle consider antiferromagnets on
a $L \times L$ square lattice with
an odd number $N_{\rm tot} = L^2$ of spin-half
moments. Such a system is expected to have a doublet ground state with total spin $S_{\rm tot}=1/2$. 
Focusing on the $S^z_{\rm tot}=1/2$ member $|G\rangle_{\uparrow}$ of this doublet, one could
examine the ground state spin texture defined by
$\Phi^z(\vec{r}) \equiv \langle S^z_{\vec{r}} \rangle_{\uparrow}$ (where $\langle \dots \rangle_{\uparrow}$
refers to expectation values in $|G\rangle_{\uparrow}$), and use the antiferromagnetic component of this
spin texture, defined as 
\begin{equation}
n^z = \frac{1}{N_{\rm tot}}\sum_{\vec{r}} \eta_{\vec{r}} \langle S^z_{\vec{r}} \rangle_{\uparrow},
\end{equation}
to obtain information about the antiferromagnetic ordering in
the system (here $\eta_{\vec{r}} = +1$ on the $A$ sublattice and $-1$
on the $B$ sublattice).

Clearly, $n^z$ provides a measure of antiferromagnetic order that
is quite distinct from the conventional order parameter $m$, which can be
defined, e.g., according to
\begin{equation}
m^2 = \frac{1}{N_{\rm tot}}\sum_{\vec{r} \vec{r}^{'}} \eta_{\vec{r}} \eta_{\vec{r}^{'}}\langle \vec{S}_{\vec{r}} \cdot \vec{S}_{\vec{r}^{'}}\rangle_{0},
\end{equation}
where $\langle\dots \rangle_{0}$ denotes averages in the singlet ground state realized for even $N_{\rm tot}$. The relationship between the
thermodynamic-limit values of $n^z$ and $m$ is a fundamental aspect of the spontaneously broken $SU(2)$ symmetry of the N\'eel state. However, not much is 
known about it beyond the fact that $n^z$ is {\em significantly smaller than} $m$ for the nearest neighbour Heisenberg antiferromagnet on ths square 
lattice.\cite{Hoglund_thesis} Here, we provide a more detailed characterization of this relationship.

Our basic result is that $n^z$ is determined in a universal way by the value of $m$.
In other words, $n^z$ plotted against $m$ for several different
deformations of the $S=1/2$ square lattice Heisenberg antiferromagnet
falls on a single curve which defines
a universal function that is insensitive to the
microscopic details of the model Hamiltonian.
This universal function is well-approximated
by a polynomial interpolation formula: 
\begin{equation}
n^z \approx (\frac{1}{3} - \frac{a}{2} -\frac{b}{4}) m + am^2+bm^3,
\end{equation}
with $a \approx 0.288$ and $b\approx -0.306$. In addition, we also
find that the full spin texture $\Phi^z(\vec{r})$ is dominated
by Fourier modes near the antiferromagnetic wave-vector
in a universal way independent of microscopic details. We show
that this universality is captured by spin-wave theory, which
also predicts 
\begin{equation}
n^z = (1-\alpha -\beta/S)m + (\alpha/S)m^2 +{\mathcal O}(S^{-2}),
\end{equation}
with $\alpha \approx 0.013$ and $\beta \approx 1.003$ for spin-$S$ antiferromagnets. In addition, we explore
two other ways of thinking about this universal function. One of
them is a mean field theory formulated in terms of the total spin
of each sublattice, while the other approach
is in terms of a quantum rotor Hamiltonian for the N\'eel vector $\vec{n}$ of
a system with an odd number of sites. Both these
give 
\begin{equation}
n^z = \frac{m}{3}
\label{onethird}
\end{equation}
 for $S=1/2$ antiferromagnets, which is close
to the observed relationship but not exactly right.  We argue that this latter
estimate  (Eqn.~\ref{onethird}) will become asymptotically exact in the limit
of infinitely long-range {\em unfrustrated} exchange interactions. In this
limit, we also expect $m \rightarrow 1/2$, and our polynomial
fit to the universal function $n^z(m)$ was therefore constrained to ensure that 
$n^z \rightarrow m/3$ when $m \rightarrow 1/2$.

The outline of the rest of the paper is as follows: In Section~\ref{models}
we define various deformations of the square lattice $S=1/2$ Heisenberg
antiferromagnet. In Section~\ref{numerics}, we outline the projector quantum Monte Carlo (QMC) method used in this study,
and then discuss in some detail our QMC results
for $n^z$ as well as the full spin texture $\Phi^z(\vec{r})$, focusing
on the universal properties alluded to earlier. 
In Section~\ref{spinwave}, we outline three analytical
approaches to the relationship between $n^z$ and $m$.
The first is a large-$S$ spinwave expansion,
within which we calculate the ground state
spin texture $\Phi^z(\vec{r})$ and its antiferromagnetic Fourier
component $n^z$ to
leading ${\mathcal O}(1/S)$ order,
and demonstrate that such a calculation also yields the universality
properties summarized earlier, but
does not provide a quantitatively accurate account of the QMC results
for $\Phi^z(\vec{r})$ or $n^z(m)$.  The second is
a mean-field theory formulated in terms of the total spin of
each sublattice. And the third approach is in terms of a quantum
rotor Hamiltonian which is expected to correctly describe the low-energy
tower of states for odd $N_{\rm tot}$.
In Section~\ref{discussion}, we conclude with
some speculations about a possible effective field theory approach
to the calculation of $\Phi^z(\vec{r})$.

\section{Models}
\label{models}

We consider four deformations of the square lattice $S=1/2$ nearest neighbour Heisenberg antiferromagnet; all four retain 
the full $SU(2)$ spin rotation symmetry of the original model. 

\begin{figure}
{\includegraphics[width=.8\linewidth]{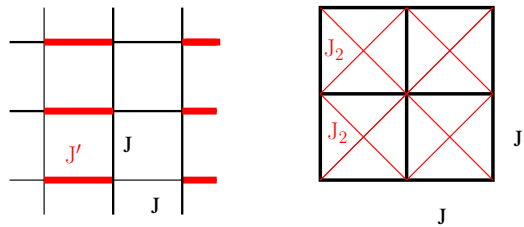}}
\caption{An illustration of the interactions present in the $JJ^{\prime}$ (left panel) and $JJ_2$ (right panel) 
model Hamiltonians. In this illustration, black bonds denote exchange interaction strength of $J$, while 
a red bond represents exchange strength of $J^{\prime}$ ($J_2$) in the left (right) panel   }
\label{jjmodels}
\end{figure}
\begin{figure}
{\includegraphics[width=.8 \linewidth]{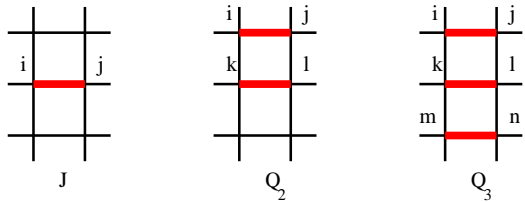}}
\caption{ Bond and plaquette operators in $JQ$ model Hamiltonians.
A thick bond denotes a bipartrite projector acting on that bond. 
All possible orientations of these bond and plaquette operators are allowed. 
}
\label{Pijkl}
\end{figure}
The first of these models is the coupled-dimer antiferromagnet, in which
there are two kinds of nearest neighbour interactions $J$ and $J'$,
as shown in Fig.~\ref{jjmodels} (left panel), where the ratio $\alpha = J'/J$ can be tuned
from $\alpha=1$ to $\alpha = \alpha_c \approx 1.90$ at which collinear
antiferromagnetic order is lost.\cite{Wenzel_Janke} The Hamiltonian for this system reads:
\be
H_{JJ'} =J \sum_{\langle ij \rangle} \textbf{S}_{i} \cdot \textbf{S}_{j}
+J' \sum_{\langle ij \rangle^{'}} \textbf{S}_{i} \cdot \textbf{S}_{j},
\label{stripedhamiltonian} 
\ee
where $\langle ij \rangle$ $(\langle ij \rangle^{'})$ denotes a pair of nearest
neighbour sites 
connected by a black (red) bond (see Fig.~\ref{jjmodels}). Another deformation of the Heisenberg model,
the $JJ_2$ model, has additional next nearest neighbour Heisenberg exchange interactions $J_2$, as shown in 
Fig.~\ref{jjmodels} (right panel). The Hamiltonian reads
\be
H_{JJ_2} =  J \sum_{\langle ij \rangle} \textbf{S}_{i} \cdot \textbf{S}_{j}+
J_2 \sum_{\langle \langle ij \rangle \rangle} \textbf{S}_{i} \cdot \textbf{S}_{j},
\label{nnnhamiltonian}
\ee
where ${\langle \langle ij \rangle \rangle}$ denotes a pair of next nearest neighbour sites.
Both these are amenable to straightforward spin-wave theory analyses, and the coupled
dimer model can also be studied numerically to obtain numerically exact results
even for very large sizes due to the
absence of any sign problems in Quantum Monte Carlo studies. However, exact numerical
results on the $JJ_2$ model are restricted to small sizes since Quantum Monte Carlo
methods encounter a sign problem when dealing with next-nearest neighbour interactions
on the square lattice. 

In addition, we study two generalizations that involve additional multispin interactions; the ``$JQ$'' 
models.\cite{Sandvik,Lou_Sandvik_Kawashima} Of these, the $JQ_2$ model has $4$-spin interactions in addition to 
the usual Heisenberg exchange terms, and is defined by the Hamiltonian
\be
H_{JQ_2} =  -J \sum_{\langle ij \rangle} P_{ij} - Q_{2} \sum_{\langle ij, kl \rangle} P_{ij}P_{kl},
\label{JQ2hamiltonian}
\ee
where the plaquette interaction $Q_2$ involves two adjacent parallel bonds
on the square lattice as shown in Fig.~\ref{Pijkl} (middle panel) and 
\begin{equation}
P_{ij} = \frac{1}{4} - {\mathbf S}_{i} \cdot {\mathbf S}_{j}
\end{equation}
is a bipartite singlet projector. The first term in Eqn. \ref{JQ2hamiltonian} is just
the standard Heisenberg exchange. Similarly, the $JQ_3$ model has $6$-spin interactions and
is defined by the Hamiltonian
\be
H_{JQ_3} = -J \sum_{\langle ij \rangle} P_{ij} - Q_{3} \sum_{\langle ij, kl, nm \rangle} P_{ij}P_{kl}P_{nm},
\label{JQ3hamiltonian}
\ee
where the plaquette interactions now involve three adjacent parallel bonds on
the square lattice, as shown in Fig.~\ref{Pijkl} (right panel). The products of singlet projectors making
up the $Q_2$ and $Q_3$ terms tend to reduce the N\'eel order of the ground state, and, when sufficiently strong,
lead to a quantum phase transition into a valence-bond-solid state.\cite{Sandvik,Lou_Sandvik_Kawashima} Here we
stay within the N\'eel state in both models, and study universal aspects
of this state as the N\'eel order is weakened.

\begin{figure}
{\includegraphics[width=\columnwidth]{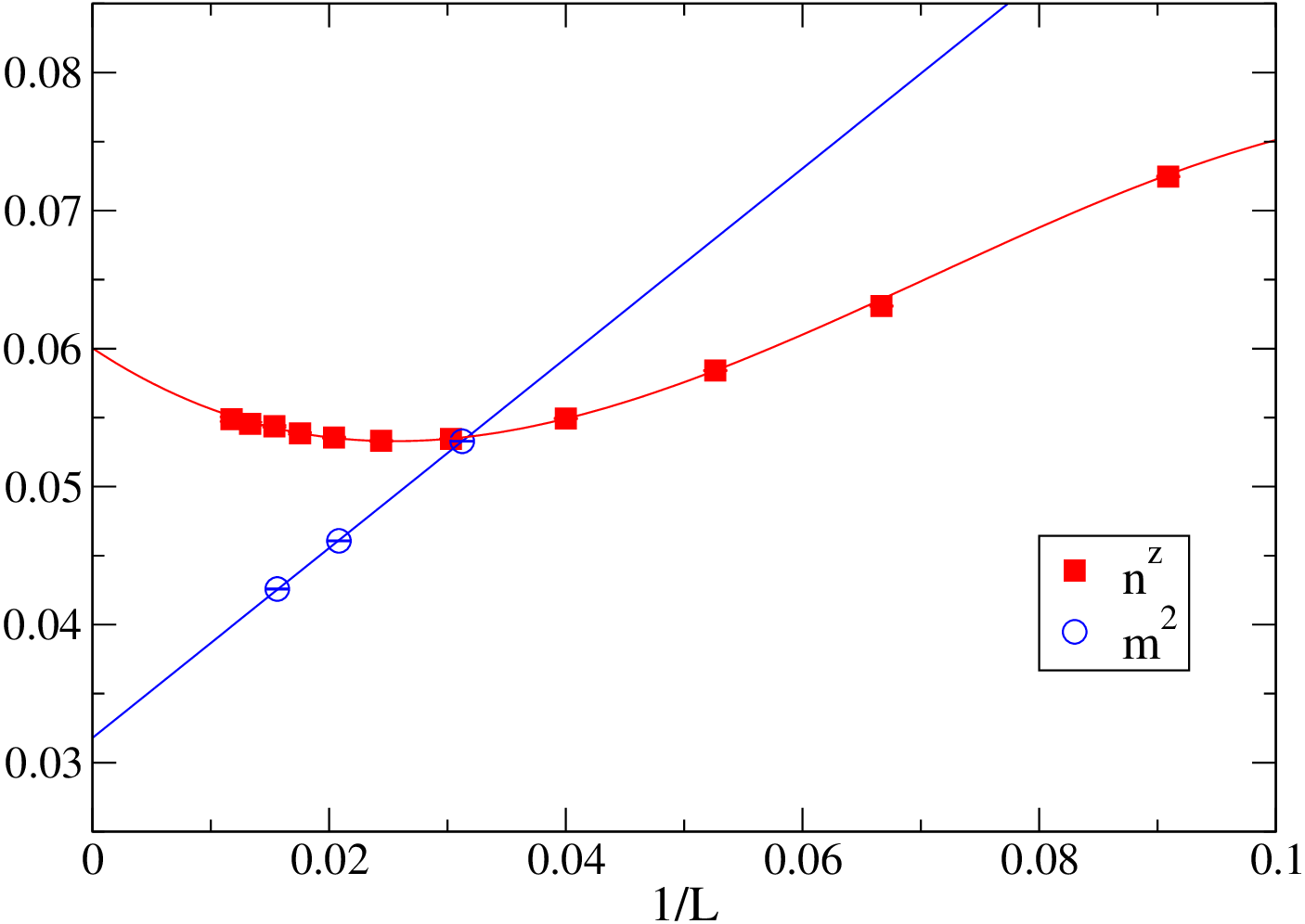}}
\caption{An illustrative example of  finite size corrections of $n^z$ and $m^2$, observed in the 
antiferromagnetic phase of the $JJ^{\prime}$ model($J^{\prime} = 1.8$). 
Note the non-monotonic behaviour of finite size corrections for 
 $n^z$, which is fitted to a  cubic polynomial. 
In contrast, finite size data for $m^2$ is well described by a  linear dependence on $1/L$.}
\label{szpp_nonmonotonic_jj}
\end{figure}

\section{Projector QMC studies}
\label{numerics}

We use the total spin-half sector version~\cite{Banerjee_Damle} of the valence-bond basis projector 
QMC method~\cite{Sandvik05,Sandvik_Evertz} to study $L \times L$ samples with $L$ odd and free boundary
conditions. We compute $\Phi^z(\vec{r})$ and $n^z$ in such
samples for the $JJ^{\prime}$ model and $JQ$ models in their antiferromagnetic phase.
We also study  the same models on $L \times L$ lattices
with $L$ even and periodic bondary conditions using the
original singlet sector valence bond projector QMC method. 
In both cases we use the most recent formulation with very efficient 
loop updates.\cite{Sandvik_Evertz,Banerjee_Damle}
Our system sizes range from $L=11$ to $L=101$, and projection power scales as
$L^3$ to ensure convergence to the ground state. We perform  $\gtrsim10^5$ equilibration 
steps followed by $\gtrsim10^6$ Monte Carlo measurements to ensure that
statistical and systematic errors are small.

\begin{figure}
{\includegraphics[width=\columnwidth]{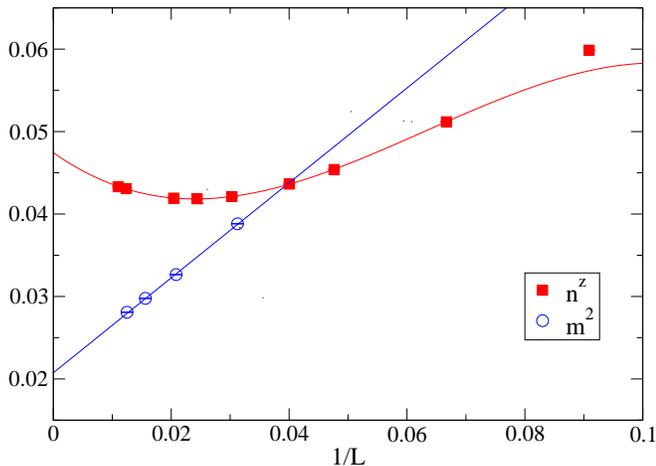}}
\caption{Another illustrative example of  finite size corrections of $n^z$ and $m^2$, observed in the 
antiferromagnetic phase of $JQ_2$ model at $Q_2 = 1.0$. Again, note the non-monotonic behaviour of finite size corrections for 
 $n^z$, which is fitted to a  cubic polynomial (only $L>20$ data used
in the fit). 
In contrast, finite size data for $m^2$ is well described by a  linear dependence on $1/L$.}
\label{szpp_nonmonotonic_jq2}
\end{figure}

Data for  $n^z$ from a sequence of $L\times L$ systems with $L$ odd
shows that  $n^z$ extrapolates to a finite value in the $L\rightarrow\infty$ limit as long as the system is in the antiferromagnetic phase.
However, we find that the approach of this observable to the thermodynamic limit has a non-monotonic behaviour. To obtain accurate 
extrapolations to infinite size, it is therefore necessary to fit the finite size data to a third-order polynomial in $1/L$. We find that the coefficient
for the leading $1/L$ term in this polynomial is rather small; this
is true for all the models studied here, as long as they 
remain in the  antiferromagnetic phase. In  Fig..~\ref{szpp_nonmonotonic_jj} and Fig.~\ref{szpp_nonmonotonic_jq2}, we show examples of this 
behaviour of the finite size corrections in $n^z$. In these figures, we also show the approach to the thermodynamic limit for $m$, as measured in a 
sequence of periodic $L \times L$ systems with $L$ even. We find that in complete contrast to the behaviour of $n^z$, $m$ extrapoloates
monotonically to the thermodynamic limit, with a dominant $1/L$ dependence---this is consistent with previous studies of the structure factor
in square lattice antiferromagnets \cite{Sandvik_Evertz} (however, with spatially anisotropic couplings, one can also observe strong 
non-monotonicity in $m$\cite{Sandvik99}).

\begin{figure}
{\includegraphics[width=\columnwidth]{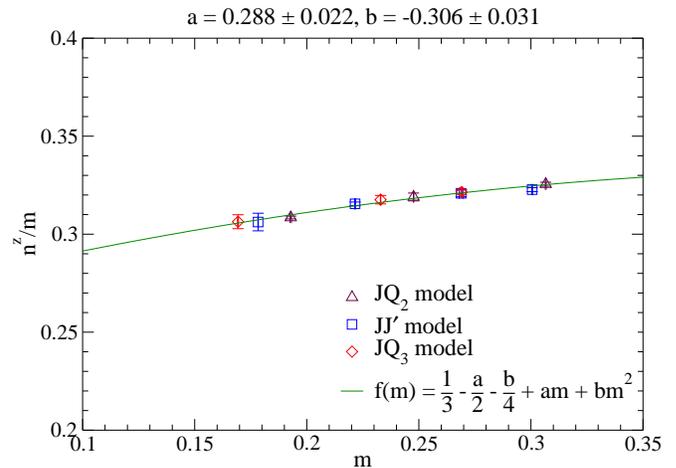}}
\caption{Extrapolated thermodynamic values of $n^z$ for three different models of 
antiferromagnets on an open  lattice, plotted as function of staggered magnetisation $m$ for the same models on periodic lattices. The former is clearly an universal function 
of the later.  This universal function can be well approximated by a polynomial fit constrained 
to ensure that $n^z(m) \rightarrow m/3$ in the limit of $m \rightarrow \frac{1}{2}$: 
$n^z \approx (1/3 - a/2 -b/4) m + am^2+bm^3$, with $a \approx 0.288$ and $b\approx -0.306$.}
\label{ms_szpp}
\end{figure}

The non-zero value of $n^z$ in the thermodynamic limit clearly reflects the long-range antiferromagnetic order present in the system and
a partial breaking of the SU$(2$) symmetry (due to the fact that we study only one member of the doublet ground state). For periodic systems, the same long range 
antiferromagnetic order is captured by the non-zero value of $m$ in the large $L$ limit---and a calculation of $m$ (through $\langle m^2\rangle$) 
for the odd-$L$ systems with periodic boundaries would of course lead to the same value. However, since $m \not= n^z$, the full staggered magnetization 
is not forced to lie along the $z$ spin axis, and it is interesting to ask: What is the relationship between these two measures of antiferromagnetic 
order? Our numerical data are unequivocal as far as this relationship is concerned, as is clear from Fig.~\ref{ms_szpp}, which shows a plot of $n^z$ versus $m$ in
the thermodynamic limit of the $JJ'$, $JQ_2$ and $JQ_3$ models. Here {\em each} point represents the result of a careful extrapolation similar to 
the examples shown in Fig.~\ref{szpp_nonmonotonic_jj} and Fig.~\ref{szpp_nonmonotonic_jq2}, and provides an accurate estimate of the corresponding 
thermodynamic limits for $n^z$ and $m$. From this figure, it is clear that $n^z$ is a universal function of $m$ independent of the microscopic structure
of the Hamiltonian. To model this universal function, we use a polynomial fit that is constrained to ensure that 
$n^z \rightarrow \frac{m}{3}$ when $m \rightarrow \frac{1}{2}$; the rationale for this constraint
will become clear in Sec.~\ref{spinwave}.
We find (Fig.~\ref{ms_szpp}) that the QMC results for $n^z(m)$ are fit well by the following functional form:
\begin{equation}
n^z(m)  =  (\frac{1}{3} - \frac{a}{2} -\frac{b}{4}) m + am^2+bm^3 \; , 
\end{equation}
with $a \approx 0.288$ and $b\approx -0.306$.

\begin{figure*}
{\includegraphics[width=2\columnwidth]{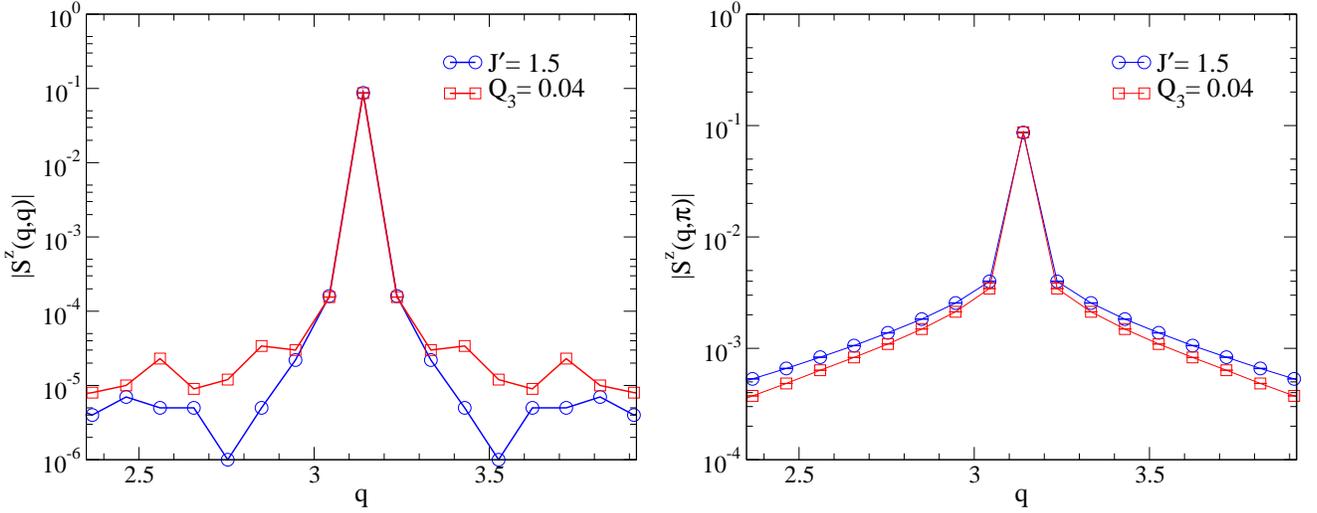}}
\caption{Fourier transform (with antiperiodic boundary conditions assumed
for convenience) of the numerically computed (for $JJ^{\prime}$ and $JQ_3$ model with $L=65$, $S=1/2$  ) $\Phi^z(\vec{r})$ along
cuts passing through the antiferromagnetic
wavevector $(\pi,\pi)$. Note the universality of the results
in the neighbourhood of the antiferromagnetic wavevector, which
in any case accounts for most of the weight in Fourier space. }
\label{texture_universality_kspace}
\end{figure*}

If one views this universal relationship as being a property
of the low energy effective field theory of the antiferromagnetic
phase, one is led to expect that the full spatial structure
of the spin texture $\Phi^z(\vec{r})$ should also be universal.
More precisely, one is led to expect that this texture
is dominated in a universal way by Fourier components near
the antiferromagnetic wavevector.
To test this, we compare the spin texture in the $JJ'$ model
and the $JQ_3$ model, choosing the strengths of
the $J'$ interaction and the $Q_3$ interaction so that
both have the same value of $m$, and therefore the same value
of $n^z$. This is shown in Fig.~\ref{texture_universality_kspace},
which shows that these very different microscopic
Hamiltonians have spin-textures whose Fourier transform
falls on top of each other at and around the antiferromagnetic wavevector.

\section{Analytical approximations}
\label{spinwave}

We now present three distinct analytical approaches to understanding
these numerical results presented in the previous section: First, we develop a 
spin-wave expansion that becomes asymptotically exact for large $S$\cite{Anderson}. 
Second, we explore a mean-field theory written in terms of the total
spin of each sub-lattice. Finally, we describe
an alternative approach in which the low-energy antiferromagnetic
tower of states of a spin-$1/2$ antiferromagnet is described by a phenomenological 
rotor model\cite{Chandrasekharan} adapted to the case of  a system with odd $N_{\rm tot}$.

\subsection{Spin-wave expansion}
\label{spinwave:1}

The leading order spin-wave calculation proceeds as usual by using
an approximate representation of spin operators in terms
of Holstein-Primakoff bosons. The resulting bosonic
Hamiltonian is truncated to leading (quadratic) order in boson operators
to obtain the first quantum corrections to the classical energy of
the system.

As is standard in the spin wave theory
of N\'eel ordered states, we start with the classical
N\'eel ordered configuration with the N\'eel vector pointing along the $\hat{z}$ axis, which corresponds to $S^z_{\vec{r}}= \eta_{\vec{r}}S$.
We then represent the spin operators at a site $\vec{r}$ of 
the square lattice in terms
of canonical bosons to leading order in $S$ as follows:
For sites $\vec{r}$ belonging to the $A$ sublattice we write
\begin{equation}
  S^{+}_{\vec{r}}=\sqrt{2S}b_{\vec{r}}  \; ;~~
  S^{z}_{\vec{r}}=S- b^{\dagger}_{\vec{r}}b_{\vec{r}} \; ,
\label{hpasublatA}
\end{equation}
while on sites $\vec{r}$ belonging to the $B$ sublattice we write
\begin{equation}
S^{-}_{\vec{r}}=\sqrt{2S}b_{\vec{r}}  \; ;~~
S^{z}_{\vec{r}}=-S+b^{\dagger}_{\vec{r}}b_{\vec{r}} \; .
\label{hpasublatB}
\end{equation}
The number of bosons at each site thus represents the effect
of quantum fluctuations away from the classical N\'eel ordered
configuration.

To quadratic order in the boson operators, this expansion yields
the following
spin wave Hamiltonian in the general case (with arbitrary two-spin
exchange couplings):
\begin{eqnarray}
 H_{sw}&=&\epsilon_{cl}S^2 + \frac{S}{2}{\mathbf{b}}^{\dagger}M{\mathbf{b}} \; , \; {\mathrm{with}} \nonumber \\
&& M_{\vec{r}\vec{r}'} = \left( \begin{smallmatrix} A_{\vec{r}\vec{r}'}&B_{\vec{r}\vec{r}'}\\ B_{\vec{r}\vec{r}'}&A_{\vec{r}\vec{r}'} \end{smallmatrix} \right) \nonumber \\
&& {\mathbf{b}_{\vec{r}}} = \left( \begin{smallmatrix} b_{\vec{r}}\\ b^{\dagger}_{\vec{r}} \end{smallmatrix} \right) \; . 
\label{swHamiltonian}
\end{eqnarray}
Here $\epsilon_{cl}S^2$ is the classical energy of the N\'eel state, $M$ in the first line is a $2N_{\rm tot}$ dimensional matrix specified
in terms of $N_{\rm tot}$ dimensional blocks $A$ and $B$, and ${\mathbf{b}}$ is
a $2N_{\rm tot}$ dimensional column vector as indicated above.
Elements of $A$ and $B$ can be written explicitly as 
\begin{eqnarray}
A_{\vec{r}\vec{r}'} & = & (Z^{U}_{\vec{r}} -Z^{F}_{\vec{r}})\delta_{\vec{r}\vec{r}'} + J^{F}_{\vec{r}\vec{r}'},\\
B_{\vec{r}\vec{r}'} & = & J^{U}_{\vec{r}\vec{r}'}.
\end{eqnarray}
In the above, $J^{F}_{\vec{r}\vec{r}'}$ are Heisenberg exchange couplings
between two sites $\vec{r}$ and $\vec{r}'$ belonging to the same
sub-lattice, $J^{U}_{\vec{r}\vec{r}'}$ are the
Heisenberg exchange couplings between sites belonging to different sublattices, 
and
\begin{eqnarray}
Z^{U}_{\vec{r}} & = & \sum_{\vec{r}'}J^{U}_{\vec{r}\vec{r}'},\\
Z^{F}_{\vec{r}} & = & \sum_{\vec{r}'} J^{F}_{\vec{r}\vec{r}'}.
\end{eqnarray}

The effects of quantum fluctuations on the classical N\'eel state can now
be calculated by diagonalizing this Hamiltonian by a canonical Bogoliubov
transformation $\mathbb{S}$ which
relates the Holstein-Primakoff bosons $b$ to the bosonic
operators $\gamma$ corresponding to spin-wave eigenstates
\begin{equation}
 {\mathbf{b}}=\mathbb{S}\Gamma, \;~~~~ \Gamma_{\mu} = \left(\begin{matrix} \gamma_{\mu} \\ \gamma^{\dagger}_{\mu}  \end{matrix} \right) \;,
\label{transformation} 
\end{equation}
where ${\mathbb{S}}$ is a $2N_{\rm tot}$ dimensional matrix that transforms
from $\mathbf{b}$ which creates and destroys bosons at specific lattice sites
$\vec{r}$ to $\Gamma$ which creates and destroys spin-wave quanta in specific spin-wave modes $\mu$.
Naturally, we must require that $H_{sw}$ be {\em diagonal} in this
new basis. We represent this diagonal form as
\begin{equation}
 H_{sw}= \epsilon_{cl}S^2+\frac{S}{2}\Gamma^{\dagger} D \Gamma \; ,
\label{diagonalform}
\end{equation}
where 
\begin{equation}
D = \left( \begin{matrix} \Lambda & 0 \\ 0 & \Lambda \end{matrix} \right),
\end{equation}
 with $\Lambda$ denoting the diagonal matrix with 
the $N_{\rm tot}$ positive spin wave frequencies $\lambda_{\mu}$ on its diagonal.

To construct a ${\mathbb{S}}$ that diagonalizes $H_{sw}$ in the $\Gamma$ basis, we look for $2N_{\rm tot}$ dimensional column vectors 
\begin{equation}
y^{\mu} = \left( \begin{matrix} u^{\mu} \\ v^{\mu} \end{matrix} \right), 
\end{equation}
which satisfy the equation
\begin{equation}
M y^{\mu}= \epsilon_\mu {\cal I} y^{\mu}
\label{paravalue}
\end{equation}
with {\em positive} values of $\epsilon_{\mu}$ equal to the positive
spin-wave frequencies $\lambda_{\mu}$ for $\mu=1,2,3 \dots N_{\rm tot}$.
Here $u^{\mu}$ and $v^{\mu}$ are $N_{\rm tot}$ dimensional vectors,
\begin{equation}
{\mathcal I}=\left( \begin{matrix} {\mathbf{1}}&{\mathbf 0}\\ {\mathbf 0}&{\mathbf{-1}} \end{matrix} \right),
\end{equation}
and ${\mathbf{1}}$ is the $N_{\rm tot}\times N_{\rm tot}$ identity matrix.
With these $y^{\mu}$ in hand, one may obtain $N_{\rm tot}$ additional solutions 
to Eqn.~\ref{paravalue}, this time with {\em negative}  $\epsilon_{N_{\rm tot}+\mu} = -\lambda_{\mu}$ 
by interchanging the roles
of the $N_{\rm tot}$ dimensional vectors $u_{\mu}$ and $v_{\mu}$ in this
construction. In other words, we have
\begin{equation}
y^{N_{\rm tot}+\mu} = \left( \begin{matrix} v^{\mu} \\ u^{\mu} \end{matrix} \right),
\end{equation}
with $\mu=1,2,3 \dots N_{\rm tot}$.

We now construct ${\mathbb{S}}$ by using these $y^{\mu}$ (with $\mu=1,2,3 \dots 2N_{\rm tot}$) as its $2N_{\rm tot}$ columns:
\begin{equation}
{\mathbb{S}} = \left(y^{1},y^{2},y^{3} \dots y^{2N_{\rm tot}} \right) \; .
\end{equation}
Clearly, this choice of ${\mathbb{S}}$ satisfies
the equation
\begin{equation}
M \mathbb{S} = {\mathcal I} \mathbb{S} {\mathcal I}D
\label{eqnsatisfiedbyS}
\end{equation}
Furthermore,
the requirement that the Bogoliubov transformed operators
$\gamma$ obey
the same canonical bosonic commutation relations as the $b$ operators implies
that  $\mathbb{S}$ must satisfy
\begin{equation}
 \mathbb{S}^{\dagger} {\cal I} \mathbb{S} = {\cal I} \; ,
\label{sproperty}
\end{equation}
This constraint is
equivalent to ``symplectic'' orthonormalization conditions:
\begin{eqnarray}
 (u^{\mu})^{\dagger} u^{\nu}-(v^{\mu})^{\dagger} v^{\nu}=\delta_{\mu\nu} \; ,\\ \nonumber 
 (u^{\mu})^{\dagger} v^{\nu}-(v^{\mu})^{\dagger} u^{\nu}=0 \; ,
\label{normalization}
\end{eqnarray}
for $\mu, \nu = 1,2,3 \dots N_{\rm tot}$.
It is now easy to see that
Eqn~\ref{eqnsatisfiedbyS} and Eqn~\ref{sproperty} guarantee that
$H_{sw}$ is indeed diagonal in the new basis, since 
\begin{equation}
{\mathbf{b}}^{\dagger}M {\mathbf{b}}=\Gamma^{\dagger} \mathbb{S}^{\dagger} M \mathbb{S} \Gamma = \Gamma^{\dagger} \mathbb{S}^{\dagger}{\mathcal I} \mathbb{S} {\mathcal I}D \Gamma = \Gamma^{\dagger} D \Gamma \; .
\end{equation}

For periodic samples, it is possible to exploit the translational
invariance of the problem and work in Fourier space to obtain
these spin-wave modes and their wavefunctions and calculate $m = S - \Delta^{'}$
correct to leading order in the spin-wave expansion---as these results
are standard and well-known,\cite{spinwaveexpansion2} we do not provide further details here. On the other hand, the corresponding
results for $L \times L$ samples with free boundary conditions and $N_A = N_B+1$ do not seem to be available in the literature, and our discussion
below focuses on this case.

We begin by noting that the non-zero entries in $A$ only connect two sites belonging to the same sublattice, 
while those in $B$ always connect sites belonging to opposite sublattices. As a result of this, the solutions 
to the equation for $y^{\mu}$ can also be expressed in terms of a single function $f_{\mu}(\vec{r})$
defined on sites of the lattice. To see this, we consider
an auxillary problem of finding $\tilde{\epsilon}_{\mu}$ such that the operator $A-B - \tilde{\epsilon}_{\mu}\eta_{\vec{r}}$
has a zero mode $f_{\mu}(\vec{r})$ (as before, $\eta_{\vec{r}}$ is $+1$ for
sites belonging to the $A$ sublattice, and $-1$ for sites belonging
to the $B$ sublattice).
 
This auxillary problem has $N_{\rm tot}$ solutions corresponding to the $N_{\rm tot}$ roots $\tilde{\epsilon}_{\mu}$ of
the polynomial equation $\text{det}(A-B-\tilde{\epsilon}_{\mu} \eta_{\vec{r}}) = 0$;
these $\tilde{\epsilon}_{\mu}$ can be of either sign. To
make the correspondence with the {\em positive} $\epsilon_{\mu}$
solutions $(u^{\mu},v^{\mu})$ (with $\mu = 1,2...N_{\rm tot}$) of the original equation
$M y^{\mu} = \epsilon_{\mu} {\mathcal I}y^\mu$, we now note
that 
\begin{equation}
\langle f_{\mu} | A-B | f_{\mu}\rangle = \tilde{\epsilon}_{\mu}N_{\mu}
\end{equation}
where 
\begin{equation}
N_{\mu} \equiv \sum_{r_{A}} |f_{\mu}(r_{A})|^2 - \sum_{r_{B}} |f_{\mu}(r_{B})|^2. 
\end{equation}
Since $A-B$ is a positive (but not positive definite) operator, this
implies that $\tilde{\epsilon}_{\mu}$ has the same sign
as $N_{\mu}$ for all non-zero $\tilde{\epsilon}_{\mu}$.
To make the correspondence with the positive $\epsilon_{\mu} \equiv \lambda_{\mu}$ solutions
($\mu=1,2 \dots N_{{\rm{tot}}}$) of the original problem, we can therefore make the ansatz
\begin{eqnarray}
 u^{\mu}_{\vec{r}_{A}}=f_{\mu}(r_{A})/\sqrt{N_{\mu}}, u^{\mu}_{r_{B}}= 0 \\ \nonumber
 v^{\mu}_{r_{B}}=-f_{\mu}(r_{B})/\sqrt{N_{\mu}}, v^{\mu}_{r_{A}}= 0
\label{positivesimplify}
\end{eqnarray}
if $N_{\mu} > 0$, or the alternative ansatz
\begin{eqnarray}
 u^{\mu}_{r_{B}}=-f_{\mu}(r_{B})/\sqrt{-N_{\mu}}, u^{\mu}_{r_{A}}= 0  \\ \nonumber
 v^{\mu}_{r_{A}}=f_{\mu}(r_{A})/\sqrt{-N_{\mu}}, v^{\mu}_{r_{B}}= 0
\label{negativesimplify}
\end{eqnarray}
if  $N_{\mu} < 0$. Here, $r_A$ ($r_B$) denotes sites belonging to
the $A$ ($B$) sublattice of the square lattice.
This ansatz clearly ensures that the $y^{\mu}$ (with $\mu = 1,2..N_{\rm tot}$) obtained in this manner satisfy the original equation with positive $\epsilon_{\mu} \equiv \lambda_{\mu}$
and are appropriately normalized.

Atlhough this approach is not the one we use in our actual computations (see below), it provides
a useful framework within which we may discuss possible zero frequency spin-wave modes,
{\em i.e} $\lambda_{\mu_0} = 0$ for some $\mu_0$: A mode $\mu_0$ with $\lambda_{\mu_0} =0$
clearly corresponds to a putative zero eigenvalue of the operator $A-B$. From
the specific form of $A-B$ in our problem, it is clear that such
a zero eigenvalue does indeed exist, and $f_{\mu_0}(\vec{r})$, the corresponding
eigenvector of $A-B$, can be written down explicitly as
\begin{equation}
f_{\mu_0}(\vec{r}) = 1
\label{zeromodeAminusB}
\end{equation}
Since this corresponds to the root $\tilde{\epsilon}_{\mu_0} = 0$ of
the auxillary problem, it can {\em in principle} be used to obtain a {\em pair} of zero frequency
modes $\epsilon_{\mu_0}$ and $\epsilon_{\mu_0 + N_{{\rm tot}}}$ for the original problemof finding $\epsilon_{\mu}$ and $y^{\mu}$ that satisfy $M y^{\mu} = \epsilon_{\mu} {\mathcal I}y^\mu$.

However, we need to ensure that the symplectic orthonormalization conditions
(Eqn.~\ref{normalization}) are satisfied by our construction of the
corresponding $y^{\mu_0}$ and $y^{\mu_0 + N_{\rm tot}}$.
This is where the restriction to a $N_{\rm tot}=L \times L$ lattice
with $N_A=N_B+1$ enters our discussion. 
For this case, $N_{\mu_0} = N_A -N_B = 1$, and we are thus
in a position to write down properly normalized zero-mode wavefunctions:
\begin{eqnarray}
 u^{\mu_0}_{\vec{r}_{A}}=f_{\mu_0}(r_{A}), u^{\mu_0}_{r_{B}}= 0 \\ \nonumber
 v^{\mu_0}_{r_{B}}=-f_{\mu_0}(r_{B}), v^{\mu_0}_{r_{A}}= 0 \; ,
\label{zeromode1}
\end{eqnarray}
and
\begin{eqnarray}
 u^{N_{\rm tot}+\mu_0}_{r}=v^{\mu_0}_{r},\\ \nonumber
 v^{N_{\rm tot}+\mu_0}_{r}=u^{\mu_0}_{r} \; .
\label{zeromode2}
\end{eqnarray}
[Parenthetically, we note that the question of zero frequency spinwave modes for
the more familiar case with $N_A = N_B$ and periodic boundary conditions has been discussed earlier in the literature\cite{Anderson} and will not be considered here.]

Thus, the equation $M y^{\mu} = \epsilon_{\mu} {\mathcal I} y^{\mu}$
has a pair of zero modes related to
each other by interchange of the $u$ and $v$ components of the mode,
and it becomes necessary to regulate intermediate steps of the calculation with a staggered
magnetic field $\hat{z}{\epsilon_{h}}\eta_{\vec{r}}$ with infinitesimal magnitude ${\epsilon_{h}}>0$ in
the $\hat{z}$ direction. Denoting the corresponding $A$ by $A^{{\epsilon_{h}}}$, we see
that $A^{{\epsilon_{h}}}-B$ is now a positive definite operator and does not
have a zero eigenvalue. Indeed, it
is easy to see from the foregoing that the corresponding eigenvalue
now becomes non-zero, yielding a positive spin-wave frequency
$\lambda^{{\epsilon_{h}}}_{\mu_0}= N_{\rm tot}{\epsilon_{h}}$. One can 
also calculate the ${\mathcal O} ({\epsilon_{h}})$ term of 
$f^{{\epsilon_{h}}}_{\mu_0}(\vec{r})$ and check that $f^{{\epsilon_{h}}}_{\mu_0}$ tends to $f_{\mu_0}(\vec{r})$ in a non-singular way as ${\epsilon_{h}} \rightarrow 0$,
from which one can obtain the corresponding $y^{\mu_0}({\epsilon_{h}})$
analytically in this limit. Thus, the contribution
of the zero mode to all physical quantities can be obtained
in the presence of a small ${\epsilon_{h}}> 0$, and the ${\epsilon_{h}} \rightarrow 0$ limit of this contribution can then be taken smoothly 
and analytically at the end of the calculation.

In our actual calculations, we use this analytical understanding of the zero frequency
spin wave mode to
analytically obtain the properly regularized zero mode contribution
to various physical quantities, while using a
computationally convenient approach to numerically calculate the
contribution of the non-zero spin wave modes.
To do this, we rewrite Eqn.~\ref{paravalue} for $\mu= 1,2,3 \dots N_{\rm tot}$
as
\begin{eqnarray}
 (A+B) \phi^{\mu}&=&\lambda_{\mu} \psi^{\mu} \\ \nonumber
(A-B) \psi^{\mu}&=&\lambda_{\mu} \phi^{\mu}
\label{phipsi}
\end{eqnarray}
where
\begin{eqnarray}
\phi^{\mu}&=& u^{\mu}+v^{\mu} \\ \nonumber
\psi^{\mu}&=& u^{\mu}-v^{\mu} .
\label{uv}
\end{eqnarray}

This implies
\begin{eqnarray}
 (A-B)(A+B)\phi^{\mu}&=&\lambda_{\mu} (A-B) \psi^{\mu} =\lambda^{2}_{\mu} \phi^{\mu} \\ \newline
 (A+B)(A-B)\psi^{\mu}&=&\lambda_{\mu} (A+B) \phi^{\mu} =\lambda^{2}_{\mu} \psi_{\mu}
\label{shortmethod1}
\end{eqnarray}

We now decompose
\begin{equation}
A-B=K^{\dagger} K.
\label{decompose}
\end{equation}
where
\begin{equation}
K= \sqrt{\omega} U.
\label{K}
\end{equation}
with $\omega$ the diagonal matrix with diagonal entries given
by eigenvalues of the real symmetric matrix $A-B$, and $U$ the matrix
whose rows are made up of the corresponding eigenvectors.

With this decomposition, we multiply Eqn~\ref{shortmethod1} by $K$ from
the left to obtain
\begin{equation}
K(A+B)K^{\dagger}\chi^{\mu} = \lambda^{2}_{\mu} \chi^{\mu}.
\label{reduced_problem}
\end{equation}
with $\chi^{\mu} = K \psi^{\mu}$.
From the solution to this equation, we may obtain the $\phi$ as
\begin{equation}
\phi^{\mu}=(K^{\dagger})\chi^{\mu}/\lambda_{\mu}.
\label{chi}
\end{equation}
and thence obtain $\psi^{\mu}$ using Eqn~\ref{phipsi}.  In order
to ensure the correct normalization of the resulting $u^{\mu},v^{\mu}$,
we impose the normalization condition 
\begin{equation}
 (\chi^{\mu})^{\dagger}\chi^{\mu}=\lambda_{\mu}.
\label{chinorm}
\end{equation}

Thus our computational strategy consists of obtaining
eigenvalues of the symmetric operator $K
(A+B)K^{\dagger}$, and using this information
to calculate the $y^{\mu}$ and thence the Bogoliubov transform
matrix ${\mathbb{S}}$. Notwithstanding the normalization
used in Eqn~\ref{chinorm}, the zero mode with $\lambda_{\mu_0} = 0$
causes no difficulties in this approach, since we work
in practice with the projection of $K(A+B)K^{\dagger}$ in the space
orthogonal to the zero mode. This is possible because we already have
an analytic expression correct to ${\mathcal O}(\epsilon_h)$ for $y^{\mu_0} (\epsilon_h)$ and $y^{N_{\rm tot}+\mu_0} (\epsilon_h)$ corresponding to this zero mode,
and do {\em not} need to determine these two columns of ${\mathbb{S}}$ by this
computational method.

We use this procedure to calculate the zero temperature boson
density as
\begin{equation}
 \langle b_{\vec{r}}^{\dagger} b_{\vec{r}} \rangle =\lim_{\epsilon_h \rightarrow 0}\sum_{\mu=1}^{N_{\rm tot}}  \left(v^{\mu}_{\vec{r}} (\epsilon_h)\right)^2.
\label{number_boson}
\end{equation}
In this expression, one may use the numerical procedure outlined
above to obtain the contribution of
all $\mu \neq \mu_0$ {\em directly at $\epsilon_h = 0$}, while
being careful to use our analytical results for
$v^{\mu_0}({\epsilon_{h}})$ to obtain the limiting value of
the contribution from $\mu= \mu_0$. This gives
\begin{equation}
  \langle b_{\vec{r}_A}^{\dagger} b_{\vec{r}_A} \rangle =\sum_{\mu \neq \mu_0} (v^{\mu}_{\vec{r}_A})^2
\label{zeromode_splitA}
\end{equation}
\begin{equation}
  \langle b_{iB}^{\dagger} b_{iB} \rangle =1 + \sum_{\mu \neq \mu_0} (v^{\mu}_{\vec{r}_B})^2
\label{zeromode_splitB}
\end{equation}
Here, the distinction between sites on the $A$ and $B$ sublattices arises in this final result
because $\lim_{{\epsilon_{h}} \rightarrow 0} v^{\mu_0}_{\vec{r}} ({\epsilon_{h}}) = -1$
for $\vec{r}$ belonging to the $B$ sublattice, while
$\lim_{{\epsilon_{h}} \rightarrow 0} v^{\mu_0}_{\vec{r}} ({\epsilon_{h}}) = 0$
for $\vec{r}$ belonging to the $A$ sublattice.

Knowing the average boson number at each site gives us the first quantum
corrections to the ground state expectation value $\langle S^z(\vec{r}) \rangle$:
\begin{equation}
\langle S^z(\vec{r})\rangle = \eta_{\vec{r}}(S -\langle b_{\vec{r}}^{\dagger} b_{\vec{r}}\rangle)
\end{equation}
This result for the spin-wave corrections to the ground state spin texture
then allows us to write $n^z = \lim_{L \rightarrow \infty} (\sum_{\vec{r}} \eta_{\vec{r}}\langle S^z(\vec{r})\rangle)/N_{\rm tot}$ as
\begin{equation}
n^z = S - \Delta
\end{equation}
where $\Delta$ represents the leading spin-wave correction to the
classical value for $n^z$.

\begin{figure}
{\includegraphics[width=\columnwidth]{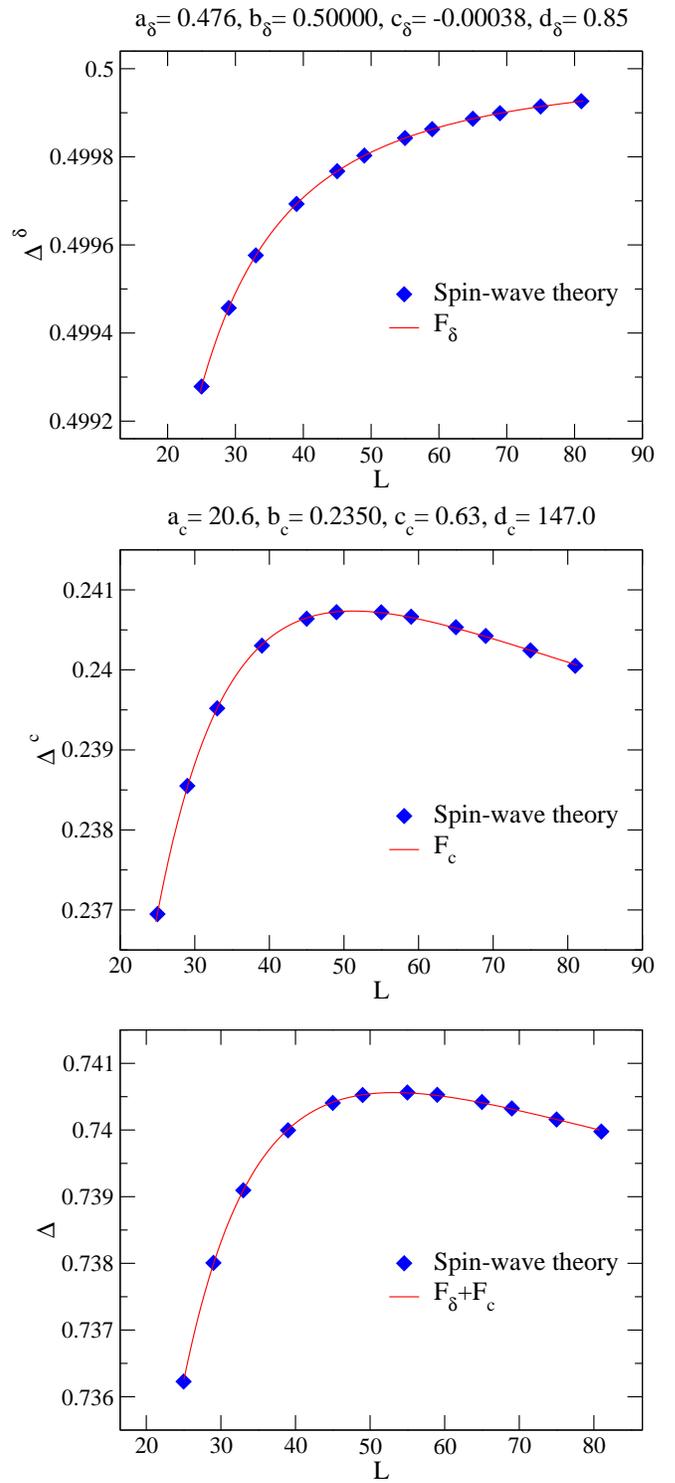}}
\caption{A typical example of the finite size scaling of the delta-function and continuum contributions to $\Delta$. Note the monotonically
increasing size dependence of the delta-function contribution,
and the non-monotonic and more slowly converging nature
of the continuum contribution. Due to this difference in their
behaviour, we find it more accurate to separately fit each of these
contributions to a polynomial
in $1/L$ and use this to obtain the thermodynamic limit of the total
$\Delta$. Here $F_{\delta/c}(L)=b_{\delta/c} +{c_{\delta/c}}/{L}-{a_{\delta/c}}/{L^2}+{d_{\delta/c}}/{L^3}$.}
\label{fullDelta_example}
\end{figure}

In order to obtain $n^z$ reliably in this manner, it is important to
understand the finite size scaling properties of $\Delta$
for various values of $J^{'}/J$ in the striped interaction model and
$J_2/J$ in the model with next-nearest neighbour interactions. In Fig.~\ref{fullDelta_example}, 
we show a typical example of this size dependence. As is clear, we find
that $\Delta$ has a non monotonic dependence on $L$: $\Delta$ initially 
increases rapidly with size, and, after a certain crossover size $L^{*}$, 
it starts decreasing slowly to finally saturate to
its asymptotic value. This non-monotonic behaviour is qualitatively
similar to that observed in the finite size extrapolations of $n^z$ from
our QMC data earlier. To explore this unusual size dependence further
and reliably extrapolate to the thermodynamic limit,
we analyze the contributions to $\Delta$ from the spin-wave spectrum
in the following way: We note that there is always
a monotonically and rapidly convergent
${\mathcal O}(1)$ contribution to $\Delta$ from the lowest
frequency spin-wave mode, whose spin-wave frequency scales to zero as $1/N_{\rm tot}$ (for any finite
$N_{\rm tot}$, this is {\em not} an exact zero mode of the system). We dub
this the `delta-function contribution' and its thermodynamic limit
is easy to reliably extrapolate to. In addition, there is a `continuum
contribution' coming from all the other spin-wave modes, each of which
contributes an amount of order ${\mathcal O}(1/N_{\rm tot})$. This contribution
converges less rapidly to the thermodynamic limit, and also happens
to be non-monotonic: it first increases quickly with increasing size,
and then starts decreasing slowly to finally saturate to the thermodynamic
limit.

The delta-function contribution can be fit best to a functional form 
\begin{equation}
F_{\delta}(L)=b_{\delta} +\dfrac{c_{\delta}}{L}-\dfrac{a_{\delta}}{L^2}+\dfrac{d_{\delta}}{L^3},
\end{equation}
with the dominant $1/L^2$ term accounting for the
monotonic increase with $L$, while the 
continuum contribution is fit to 
\begin{equation}
F_{c}(L)=b_c+\dfrac{c_c}{L}-\dfrac{a_c}{L^2} + \dfrac{d_c}{L^3}, 
\end{equation}
whereby the size dependence is predominantly determined
by the competition between the term proportional to $1/L$ which decreases
with increasing $L$, and the term proportional to $1/L^2$ which increases
with increasing $L$. This gives rise to non-monotonic behaviour whereby the continuum contribution first increases rapidly and
then decreases slowly beyond a crossover length $L^{*}$ to finally saturate to its infinite volume limit. We also find that the length $L^{*}$ gets larger
as we deform away from the pure square lattice antiferromagnet, making
it harder to obtain reliable extrapolations to the thermodynamic limit.

\begin{figure}
{\includegraphics[width=\columnwidth]{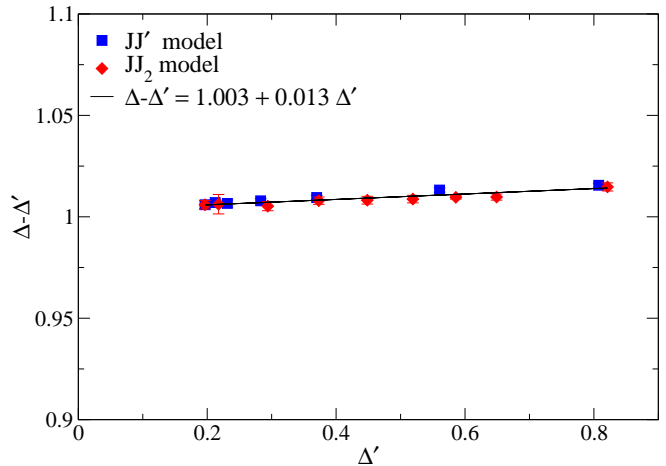}}
\caption{$\Delta -\Delta^{\prime}$, the difference
between the leading spin wave corrections to $n^z$ and $m$, plotted against
the leading spin-wave corrections $\Delta^{\prime}$ to $m$ for the $JJ^{'}$ and
$JJ_2$ models described in the text.}
\label{spinwaveuniversality}
\end{figure}

Using such careful finite-size extrapolations to obtain $\Delta$ for various
values of $J_2/J$ and $J^{'}/J$, we compare the result
with $\Delta{'}$ calculated analytically. Specifically, we now ask if the universality
seen in our QMC results is reflected in these semiclassical
spin-wave corrections to $n^{z}$ and $m$. The answer is
provided by Fig.~\ref{spinwaveuniversality}, which shows that
the numerically obtained spin-wave corrections apparently
satisfy a universal
linear relationship
\begin{equation}
\Delta - \Delta^{\prime} \approx 1.003 + 0.013\Delta^{\prime}
\end{equation}
as one deforms away
from the pure square lattice antiferromagnet in various ways.

What does this imply for $n^z(m)$ to leading order in $1/S$? To answer this, we note
that
\begin{equation}
\frac{n^z}{m} = 1 -\frac{\Delta - \Delta^{\prime}}{S} + {\mathcal O}(S^{-2})
\end{equation}
Using our numerically established universal result to relate
$\Delta - \Delta^{\prime}$ to $\Delta^{\prime}$ and thence to
$m$ itself, we obtain the universal relationship
\begin{equation}
n^z = \alpha m  + \beta m^2
\end{equation}
with $\alpha \approx 0.987 - 1.003/S$ and $\beta \approx  0.013/S$.
However, being a large-$S$ expansion, spin-wave theory is unable
to give a quantitatively correct prediction for $n^z(m)$ for the $S=1/2$
case.

\begin{figure*}
{\includegraphics[width=2\columnwidth]{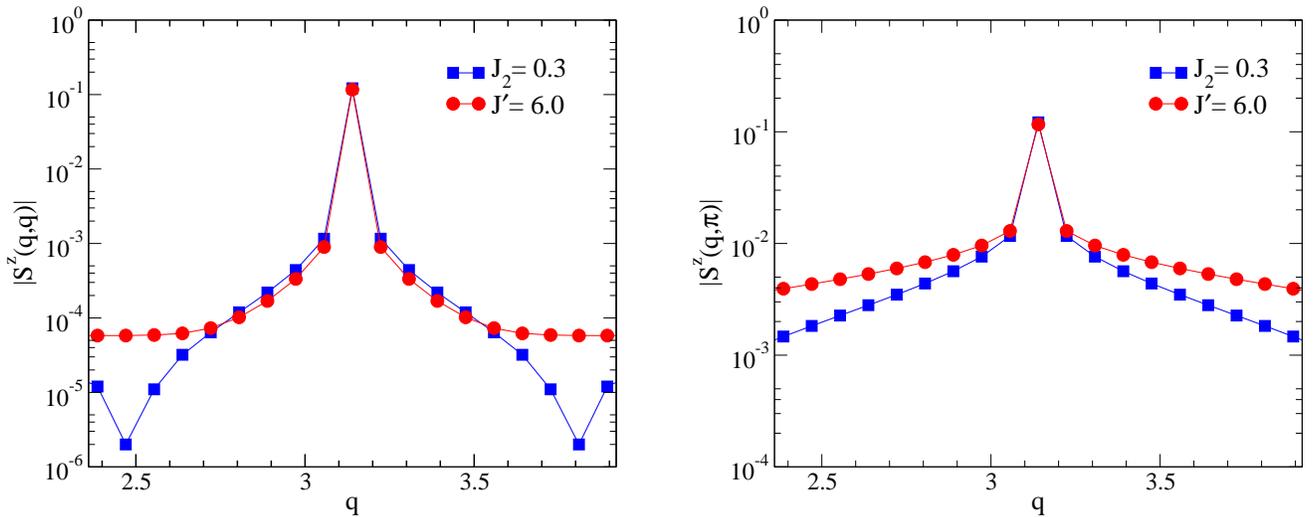}}
\caption{Fourier transform (with antiperiodic boundary conditions assumed
for convenience) of the spin-wave result for $\Phi^z(\vec{r})$ (assuming
$S=3/2$ and calculated using $L=75$ for $JJ_{2}$ and $JJ^{\prime}$ model) along
cuts passing through the antiferromagnetic
wavevector $(\pi,\pi)$. Note the nearly universal nature of the results
in the neighbourhood of the antiferromagnetic wavevector, which
in any case accounts for most of the weight of the transformed
signal.}
\label{spinwaveFTuniversality}
\end{figure*}

Finally, we use our spin-wave predictions for the ground-state spin
texture to look at the Fourier transform
of the spin-texture for various deformations of the pure
antiferromagnet. The results are shown in Fig.~\ref{spinwaveFTuniversality},
which demonstrates that spin-wave theory also predicts that the
Fourier transform of the spin-texture near the antiferromagnetic wave-vector
is a universal function of the wavevector; this provides some rationalization
for the observed universality of the Fourier transformed spin texture
seen in our QMC numerics.

\subsection{Sublattice-spin mean-field theory}
We now turn to a simple mean-field picture
in terms of the dynamics of the total spins $\vec{S}_A$
and $\vec{S}_B$ of the $A$ and $B$ sublattices respectively.
When $N_A=N_B+1$, it is clearly appropriate to assume that the
total spin quantum number of $\vec{S}_A$ is $S_B+1/2$ while the
total spin quantum number of $\vec{S}_B$ should be taken to be $S_B$,
where $S_B = N_B/2$ tends to infinity in the thermodynamic limit.

In this mean-field treatment, we assume that $\vec{S}_A$ and $\vec{S}_B$
are coupled antiferromagnetically in the effective Hamiltonian that describes the low energy part of the spectrum:
\begin{equation}
H_{MF} = J_{MF} \vec{S}_A \cdot \vec{S}_B
\end{equation}
with $J_{MF} > 0$.  Within this mean-field treatment,
the $S_{\rm tot} =1/2$, $S^z_{\rm tot} = 1/2$
ground state that we focus on in our numerics is thus the $S_{\rm tot} =1/2$,
$S^z=1/2$ state obtained by the quantum mechanical addition
of angular momenta $S_B$ and $S_B+1/2$. Within this mean-field theory,
$n^z$ is modeled as
the expectation value of $(S_A^z - S_B^z)/N_{\rm tot}$ in this state,
which can be readily obtained in closed form using the following standard result
for the minimum angular momentum state $|J=j_1-j_2, m_J\rangle$ state obtained by the addition of angular momenta $j_1$ and $j_2$ (with $j_1 \geq j_2$):
\begin{equation}
\langle j_1,m_1;j_2,m_2|J,m_J\rangle = \rho_Jc^{J,m_J}_{m_1,m_2}
\end{equation}
with
\begin{equation}
\rho_J= \sqrt{\frac{(2J+1)!(2j_2)!}{(2j_1+1)!}}
\end{equation}
and
\begin{widetext}
\begin{eqnarray}
c^{J,m_J}_{m_1,m_2} &=& (-1)^{j_2+m_2}\left[(j_1+m_1)!((j_1-m_1)!\right]^{1/2}\left[(j_2+m_2)!(j_2-m_2)!(J+m_J)!(J-m_J)!\right]^{-1/2}
\end{eqnarray}
\end{widetext}
for $m_1+m_2=m_J$ and $c^{J,m_J}_{m_1,m_2}=0$ otherwise.

In our case, $j_1=S_B+1/2$, $j_2 = S_B$, $J=1/2$, $m_J=1/2$,
and $n^z= \langle m_1-m_2\rangle_{J,m_J}/N_{\rm tot}$ can therefore
be readily calculated to obtain
\begin{equation}
n^z = \left(\frac{2}{3}S_B + \frac{1}{2}\right)/N_{\rm tot}
\end{equation}
within this phenomenological approach. 

On the other hand, when $N_A=N_B$, we may also calculate $m^2 = \langle (\vec{S}_A-\vec{S}_B)^2\rangle_{J=0}/N_{\rm tot}^2$ within the same sublattice-spin approach
\begin{equation}
m^2= (4S_B^2+4S_B)/N^2_{\rm tot} \; .
\end{equation}
This allows us to compute the ratio $n^z/m$ in
the thermodynamic limit:
\begin{equation}
n^z = \frac{1}{3}m + {\mathcal O}\left(\frac{1}{N_{\rm tot}}\right) 
\end{equation}

Is there a limit in which this sublattice-spin mean-field theory
is expected to give
exact results? To answer this, we note that the sublattice-spin
model represents the Hamiltonian of an infinite-range
model in which {\em every} $A$ sublattice-spin interacts with
{\em every} $B$ sublattice-spin  via a {\em constant}
(independent of distance) antiferromagnetic exchange coupling $J_{MF}$.
Thus, our mean-field theory is expected to become asymptotically
exact in the limit of infinitely long-range unfrustrated couplings.
In this limit, we also expect $m \rightarrow 1/2$,
and thus, our mean field theory predicts that $n^z \rightarrow m/3$
when $m \rightarrow 1/2$. This is the constraint that
we built into our choice of polynomial fit for $n^z(m)$ in Sec.~\ref{numerics}.

\subsection{Quantum rotor Hamiltonian}

When any continuous symmetry is broken, the corresponding order parameter variable becomes very ``heavy'' in a well-defined sense.\cite{Anderson} The
long-time, slow dynamics
of this heavy nearly classical variable is controlled by
an effective ``mass'' that diverges in the thermodynamic limit.

For a N\'eel ordered magnet, the order parameter is the N\'eel vector
$\vec{n}$. In the usual case of an antiferromagnet with an even
number of $S=1/2$ moments, the low-energy effective Hamiltonian
that controls the orientational dynamics of the N\'eel vector $\vec{n}$ is
\begin{equation}
H_{rotor} = \frac{\vec{L} \cdot \vec{L}}{2\chi N_{\rm tot}}
\end{equation}
where $\vec{L}$ is the angular momentum conjugate to the ``quantum rotor''
coordinate $\hat{n} \equiv \vec{n}/|\vec{n}|$,
$\chi$ is the uniform susceptibility per spin, and $N_{\rm tot}$
is the total number of spins.

What about our case with $N_A= N_B+1$ and an odd number of spins $N_{\rm tot}$?
Following earlier work on quantum rotor descriptions of insulating
antiferromagnets doped with a single mobile charge-carrier\cite{Chandrasekharan}, we postulate that the correct rotor description of our problem
is in terms of a rotor Hamiltonian in which $\vec{L}$ is
replaced by the angular momentum operator $\vec{L}^{\prime}$ conjugate
to a quantum rotor coordinate $\hat{n}$ that now
parametrizes a unit-sphere with a fundamental magnetic monopole at its
origin.\cite{Yang}
In other words, we postulate a low-energy effective Hamiltonian
\begin{equation}
H^{1/2}_{rotor} = \frac{\vec{L}^{'} \cdot \vec{L}^{'}}{2 \chi N_{\rm tot}}
\end{equation}
where the superscript reminds us that the lowest allowed angular
momentum quantum number $l$ of the modified angular momentum operator
$\vec{L}^{'}$ is $l=1/2$.

In the notation of Ref~\onlinecite{Yang}, the angular wavefunction
of the $l=1/2$, $m_l = 1/2$ ground state of this modified
rotor Hamiltonian is the {\em monopole harmonic} 
$Y_{1/2, 1/2, 1/2}(\theta, \phi)$.
To model $\langle n^z \rangle_{\uparrow}$, we must compute
the expectation value $\langle \cos(\theta) \rangle_{1/2,1/2,1/2}$
and multiply this result by $m \equiv |\vec{n}|$.
To do this we note that 
\begin{equation}
|Y_{1/2, 1/2, \pm 1/2}(\theta, \phi)|^2 = \frac{1}{4 \pi}(1 \pm \cos(\theta)) \; ,
\end{equation}
which immediately implies
\begin{equation}
\langle n^z \rangle_{\uparrow} = m\int d\cos(\theta) d \phi \cos(\theta)|Y_{1/2, 1/2, 1/2}(\theta, \phi)|^2 = \frac{1}{3} m
\end{equation}

Thus, a more general phenomenological approach that goes beyond
sublattice-spin mean-field theory but ignores all non-zero wavevector modes
also gives 
\begin{equation}
n^z = \frac{m}{3} \; .
\end{equation}
Since our QMC data show clear deviatons from this result, we
conclude that such non-zero wavevector modes are essential for
a correct calculation of the universal function $n^z(m)$.

\bigskip
\section{Discussion}
\label{discussion}
A natural question  that arises from our results is whether the universal ground state
spin texture we have found here can be successfully described using
an effective field theory approach of the type used
recently by Eggert and collaborators for studying universal
aspects of the alternating order induced by missing spins in
two dimensional $S=1/2$ antiferromagnets.\cite{Eggert}
This approach uses a non-linear sigma-model description of
the local antiferromagnetic order parameter, with lattice
scale physics only entering via the values of the stiffness
constant $\rho_s$ and the transverse susceptibility $\chi_{\perp}$,
and the presence of the vacancy captured by a local term in the action.
An analogous treatment for our situation would need two things---one
is a way of restricting attention to averages in the $S_{\rm tot}=1/2$
component $|G\rangle_{\uparrow}$ of the ground
state doublet, and the other is an understanding of the
right boundary conditions or boundary terms in the action,
so as to correctly reflect that fact that our finite sample
has open boundaries. We leave this as an interesting direction
for future work, which may shed some light on the role of
non-zero wavevector modes that were left out of the rotor
description of the earlier section.

\section{Acknowledgements}

We thank L.~Balents, A.~Chernyshev, M.~Metlitski, S.~Sachdev, R.~Shankar and R.~Loganayagam for useful discussions. The work of KD was supported 
by Grants DST-SR/S2/RJN-25/2006 and IFCPAR/CEFIPRA Project 4504-1, and that of AWS by NSF Grant No.~DMR-1104708. The numerical 
calculations were carried out using computational resources of TIFR. AWS gratefully acknowledges travel
support from the Indian Lattice Gauge Theory Initiative at TIFR.


\begin{thebibliography}{999}

\bibitem{Neuberger_Ziman} H.~Neuberger and T.~Ziman, Phys. Rev. B {\bf 39}, 2608(1989).

\bibitem{Chernyshev_White}

S.~R.~White and A.~L.~Chernyshev, Phys. Rev. Lett. {\bf 99}, 127004 (2007).

\bibitem{Sandvik_prb97}A.~W.~Sandvik, Phys. Rev. B {\bf 56}, 11678 (1997).

\bibitem{Beard_Wiese_prb}B.~B.~Beard, R.~J.~Birgeneau, M.~Greven, and U.-J. Wiese, Phys. Rev. Lett. {\bf 80}, 1742 (1998).

\bibitem{Lieb_Mattis} E. Lieb and D. C. Mattis, J. Math. Phys. {\bf 3}, 749 (1962).

\bibitem{Hoglund_thesis} K.~Hoglund Ph.D thesis (2010); K.~Hoglund and A.~W.~Sandvik, unpublished.

\bibitem{Wenzel_Janke}S. Wenzel and W. Janke, Phys. Rev. B{\bf 79}, 014410(2009).

\bibitem{Sandvik} A.~W.~Sandvik, Phys. Rev. Lett. {\bf 98}, 227202 (2007).

\bibitem{Lou_Sandvik_Kawashima}J.~Lou, A.~W.~Sandvik, and N.~Kawashima, Phys. Rev. B {\bf 80}, 180414 (2009).

\bibitem{Banerjee_Damle}A. Banerjee and K. Damle, J. Stat. Mech. (2010) P08017.

\bibitem{Sandvik05} A.~W.~Sandvik, Phys. Rev. Lett. {\bf 95}, 207203 (2005).

\bibitem{Sandvik_Evertz} A.~W.~Sandvik, and  H.~G.~Evertz, Phys. Rev. B {\bf 82}, 024407 (2010).

\bibitem{Sandvik99} A. W. Sandvik, Phys. Rev. Lett. {\bf 83}, 3069 (1999).

\bibitem{Anderson} P.~W.~Anderson, Phys. Rev. {\bf 86}, 694 (1952).

\bibitem{spinwaveexpansion2}P. Chandra and B. Doucot, Phys. Rev. B {\bf 38}, 9335, 1988

\bibitem{Colpa1} J. H. P. Colpa,  Physica A {\bf 2}, 134, 377-416  (1986); J. H. P. Colpa,  Physica A {\bf 2}, 134, 417-422  (1986)

\bibitem{Chandrasekharan} S.~Chandrasekharan, F.-J.~Jiang, M.~Pepe,
and U.-J.~Wiese, Phys. Rev. D {\bf 78}, 077901 (2008).

\bibitem{Yang} T.~T.~Wu and C.~N.~Yang, Nuc. Phys. {\bf B107}, 365 (1976);
Phys. Rev. D {\bf 16}, 1018 (1977).

\bibitem{Eggert} S. Eggert, O.F. Syliu\aa{}sen, F. Anfuso, M.Andres, Phys. Rev. Lett. {\bf 99}, 097204 (2007).










\end{thebibliography}
\end{document}